\documentclass[aps,prl,twocolumn,showpacs,groupedaddress]{revtex4-1}
\usepackage{graphicx}
\usepackage{dcolumn}
\usepackage{amsmath}
\usepackage{amsfonts}
\usepackage{amssymb}
\begin{document}
\title{Realization of giant magnetoelectricity in helimagnets}
\author{Sae Hwan Chun,$^{1}$ Yi Sheng Chai,$^{1}$ Yoon Seok Oh,$^{1}$ Deepshikha Jaiswal-Nagar,$^{1}$ So Young Haam,$^{1}$ Ingyu Kim,$^{1}$ Bumsung Lee,$^{1}$ Dong Hak Nam,$^{1}$ Kyung-Tae Ko,$^{2}$ Jae-Hoon Park,$^{2}$ Jae-Ho Chung,$^{3}$ and Kee Hoon Kim$^{1}$}

\address{$^{1}$FPRD, Department of Physics and Astronomy, Seoul National University, Seoul 151-747, South Korea\\
$^{2}$Department of Physics \& Division of Advanced Materials Science, POSTECH, Pohang 790-784, South Korea\\
$^{3}$Department of Physics, Korea University, Seoul 136-713, South Korea}

\begin{abstract}
We show that low field magnetoelectric (ME) properties of helimagnets Ba$_{0.5}$Sr$_{1.5}$Zn$_{2}$(Fe$_{1-x}$Al$_{x}$)$_{12}$O$_{22}$ can be efficiently tailored by Al-substitution level. As \emph{x} increases, the critical magnetic field for switching electric polarization is systematically reduced from $\sim$1 T down to $\sim$1 mT, and the ME susceptibility is greatly enhanced to reach a giant value of 2.0 $\times$ 10$^{4}$ ps/m at an optimum $x$ = 0.08. We find that control of nontrivial orbital moment in the octahedral Fe sites through the Al-substitution is crucial for fine tuning of magnetic anisotropy and obtaining the conspicuously improved ME characteristics.
\end{abstract}

\maketitle

In recent years, extensive researches on multiferroics have been performed with motivations to understand the nontrivial cross-coupling mechanism between magnetism and ferroelectricity as well as to search for new materials applicable in next-generation devices~\cite{Kimura01,Hur,Cheong,Ramesh,Nan,Katsura,Mostovoy,Sergienko}. Numerous studies have focused particularly on the class of so-called magnetic ferroelectrics~\cite{Cheong,Kimura01,Kenzelmann,Hur,Chapon}, in which ferroelectricity is induced by magnetic order through either inverse Dzyaloshinskii-Moriya effects~\cite{Katsura,Mostovoy,Sergienko} or the exchange striction mechanism~\cite{Chapon}. Although dramatic variation of electric polarization $P$ with magnetic field $B$, often realized in the magnetic ferroelectrics~\cite{Cheong}, might be useful for application, the phenomena occur mostly at low temperatures~\cite{Kimura01,Hur,Kenzelmann,Chapon} and related ME susceptibility is yet too small~\cite{Nan}. Hence, it is a longstanding challenge in the research of multiferroics to improve both the operating temperature \cite{Kimura02,Kimura03} and the ME sensitivity \cite{Eerenstien,Dong,Ishiwata}.

The hexaferrite Ba$_{0.5}$Sr$_{1.5}$Zn$_{2}$Fe$_{12}$O$_{22}$ (BSZFO) with helical spin order is currently a unique candidate that can show the $B$-induced ferroelectricity above room temperature up to $\thicksim$340 K~\cite{Kimura03}. However, at 300 K, its ferroelectric (FE) phase is expected to emerge at $B$ $\sim$1 T, which is too high for memory applications. Moreover, its ME coupling is rather weak; the MES $\alpha$$_{\rm ME}$ $\equiv$ $\mu_{0}$ $\frac{dP}{dB}$ at 30 K shows a maximum value of $\sim$1.3 $\times$ 10$^{3}$ ps/m at B $\sim$400 mT ~\cite{Kimura03}, which is one or two orders of magnitude smaller than the highest $\alpha$$_{\rm ME}$$\sim$10$^{4}$ - 10$^{5}$ ps/m realized in heterogeneous films ~\cite{Eerenstien} or strain-coupled composites ~\cite{Nan,Dong}. When Zn is replaced by Mg to form a Mg$_2$Y-type hexaferrite, Ba$_{2}$Mg$_{2}$Fe$_{12}$O$_{22}$, the critical magnetic field for inducing $P$ becomes extremely low, $\sim$30 mT ~\cite{Ishiwata,Taniguchi}. On the other hand, a maximum operation temperature of the Mg$_2$Y-type hexaferrite is expected to be lower than 195 K. Moreover, a microscopic understanding for the lowered critical magnetic field remains unclear. Therefore, systematic studies are further required to understand the origin of the intricate ME effect and improve multiferroic properties in the hexaferrite system.
\begin{figure}[tbp]
\begin{center}
\includegraphics[width=0.44\textwidth]{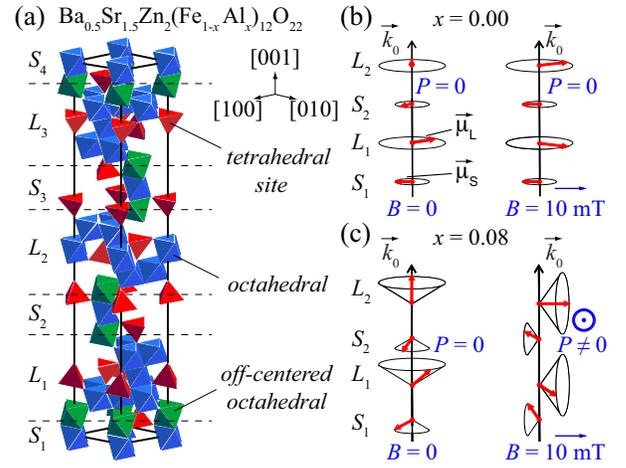}
\end{center}
\caption{(a) Crystal structure of the Zn$_{2}$Y type-hexaferrite Ba$_{0.5}$Sr$_{1.5}$Zn$_{2}$(Fe$_{1-x}$Al$_{x}$)$_{12}$O$_{22}$ that has alternating stacks of magnetic $S$ and $L$ blocks along the $c$-axis. Schematic illustration of rotating magnetic moments in the $L$ ($\vec{\mu}$$_L$) and $S$ blocks ($\vec{\mu}$$_S$) in the (b) helical ($x$ = 0.00) and (c) heliconical ($x$ = 0.08) phases under in-plane $B$ = 0 T and 10 mT. $\vec{k}$$_0$ is the spin modulation wave vector parallel to [001].} \label{FIG. 1}
\end{figure}

In this letter, we demonstrate that Al-substituted BSZFO, i.e. Ba$_{0.5}$Sr$_{1.5}$Zn$_{2}$(Fe$_{1-x}$Al$_{x}$)$_{12}$O$_{22}$ greatly improves the multiferroic characteristics, resulting in the highest $\alpha$$_{\rm ME}$ of single-phase multiferroics near-zero magnetic field. We find that predominant substitution of Al ions into octahedral Fe sites with nontrivial orbital moment is crucial for fine tuning of the magnetic anisotropy and thus for tailoring the ME coupling.

Single crystals of Ba$_{0.5}$Sr$_{1.5}$Zn$_{2}$(Fe$_{1-x}$Al$_{x}$)$_{12}$O$_{22}$ were grown from Na$_{2}$O-Fe$_{2}$O$_{3}$ flux in air ~\cite{Momozawa}. Crystals were cut into a rectangular form for electric polarization $P$  measurements along the $ab$-plane while $B$ was applied along the direction normal to the $P$ vector in the $ab$-plane. Before the ME current measurements, each specimen was electrically poled in its paraelectric state, for example, at $B$ = 0 T for $x$ $<$ 0.05 and $B$ = 2.5 T for $x$ $>$ 0.05 at 70 K, and cooled to the target temperature in the FE state (at $B$ = 1.2 T). An LCR meter was used for dielectric constant $\varepsilon$ measurements at 4 MHz. While determined FE phase boundaries were confirmed to be independent of frequency from 100 - 10 MHz below 100 K, the specific frequency was chosen because the loss was more or less minimal ($<\sim$0.3), and thus $\varepsilon$($B$) peaks could be clearly identified even up to higher temperatures.
\begin{figure}[tbp]
\centering
\includegraphics[width=0.44\textwidth]{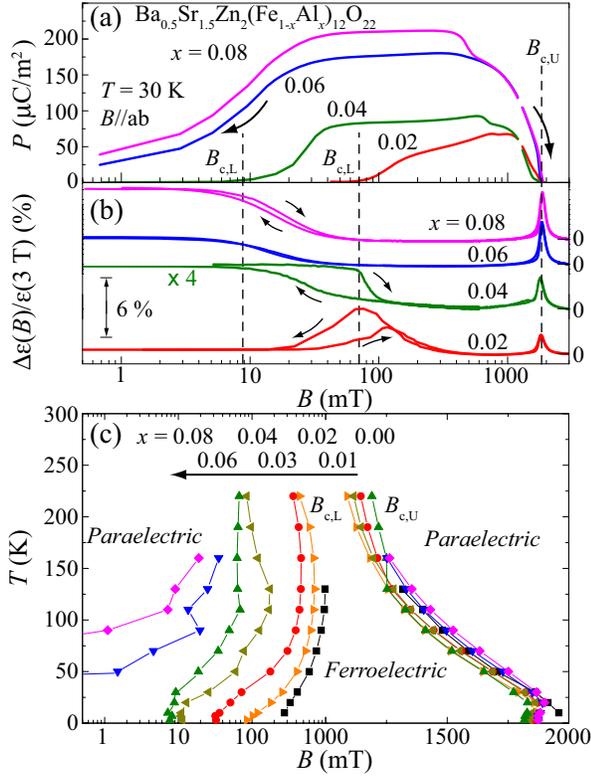}
\caption{(a) Electric polarization and (b) $\Delta$$\varepsilon$($B$)/$\varepsilon$(3 T) $\equiv$ [$\varepsilon$($B$) - $\varepsilon$(3 T)]/$\varepsilon$(3 T) curves for $x$ = 0.02, 0.04, 0.06 and 0.08 at $T$ = 30 K. (c) Ferroelectric phase boundaries determined from the $\Delta$$\varepsilon$-maxima while ramping down $B$.} \label{Fig. 2}
\end{figure}

BSZFO is one of the Y-type hexaferrites with alternating layers of tetrahedral Fe/Zn and octahedral Fe sites (Fig. 1(a)) ~\cite{Kimura03,Momozawa}. It undergoes a thermal transition from a paramagnetic to a helical spin state at $\sim$337 K, in which alternating stacks of magnetic $L$ and $S$ blocks develop a proper-screw-type rotation of their magnetic moments ($\vec{\mu}$$_L$ and $\vec{\mu}$$_S$) along [001]. In each magnetic block, the moment prefers to lie within the \emph{ab}-plane rather strongly so that the moment remains so even under an in-plane $B$ = 10 mT (modified helix phase, Fig. 1(b))\cite{Momozawa}. When $B$ is increased further, BSZFO exhibits a few successive intermediate phases (I, II, and III) and finally reaches a collinear ferrimagnetic phase around 2 T. Finite $P$ develops only in the intermediate-III phase stabilized around 1 T ~\cite{Kimura03}, of which origin remains unresolved.

When a small amount of Al is substituted for Fe in BSZFO, we find remarkable features in curves of $P$ $vs.$ $B$ and $\varepsilon$ $vs.$ $B$ (Fig. 2). All the samples yield induced $P$ in finite $B$-windows between a lower critical field ($B$$_{\rm c,L}$) and an upper critical field ($B$$_{\rm c,U}$) (Fig. 2(a)). As $x$ increases, $B$$_{\rm c,L}$ is greatly reduced, and thus a finite $P$ is observed even at $B$ $\leqq$ 1 mT for $x$ = 0.06 and 0.08. Meanwhile, the maximum of the electric polarization ($P$$_{\rm max}$) gradually increases with $x$ up to $x$ = 0.08. When $x$ is increased further, to 0.20, the turn-on behavior of $P$ becomes broader and $P$$_{\rm max}$ is substantially reduced (not shown). Figure 2(b) shows the magnetodielectric (MD) effects $\Delta$$\varepsilon$($B$)/$\varepsilon$(3 T) $\equiv$ [$\varepsilon$($B$) - $\varepsilon$(3 T)]/$\varepsilon$(3 T) of the selected samples at 30 K. The $\Delta$$\varepsilon$($B$)/$\varepsilon$(3 T) curves exhibit maxima at positions concomitant with both $B$$_{\rm c,L}$ and $B$$_{\rm c,U}$. Thus, the $\Delta$$\varepsilon$-maxima can be traced to identify the FE phase boundaries in a wide temperature-magnetic field window (Fig. 2(c)). As is evident in Fig. 2(c), $B$$_{\rm c,L}$ systematically decreases from $\sim$1 T ($x$ = 0.00) to $\sim$1 mT ($x$ = 0.08) at $T$ = 90 K, with a step of  $B$$_{c}$ $\sim$100 - 200 mT per $\Delta$$x$ = 0.01, and it becomes lower than 1 mT for $x$ = 0.08 below 90 K. These results demonstrate that Al-substitution into BSZFO is one of the most effective ways to control $B$$_{\rm c,L}$, down to 1 mT or even lower.

Another conspicuous feature of Al-substituted BSZFO is the realization of giant $\alpha$$_{\rm ME}$ at extremely low magnetic fields. Figure 3(a) shows the $B$-dependent $\alpha$$_{\rm ME}$  curves at 30 K for different $x$ and the inset summarizes their maximum values ($\alpha$$_{\rm m}$). $\alpha$$_{\rm m}$ for the undoped sample ($x$ = 0.00) is about 4.3 $\times$ 10$^{1}$ ps/m. As $x$ increases, $\alpha$$_{\rm m}$ increases to 2.0 $\times$ 10$^{4}$ ps/m at $x$ = 0.08, 470 times larger than at $x$ = 0.00, and then decreases for $x$ $>$ 0.08. Such a large enhancement in $\alpha$$_{\rm m}$ could be realized because the Al-substitution not only increases $P$ but also induces $P$ in a lower $B$-region. The present optimal $\alpha$$_{\rm m}$ = 2.0 $\times$ 10$^{4}$ ps/m is $\sim$12 times larger than $\alpha$$_{\rm m}$ $\sim$ 1.6 $\times$ 10$^{3}$ ps/m of Ba$_{2}$Mg$_{2}$Fe$_{12}$O$_{22}$ ~\cite{Ishiwata}, and records the highest value among single-phase multiferroics ~\cite{Schmid}; for comparison, $\sim$10$^{2}$ ps/m in TbMnO$_{3}$ ~\cite{Kimura01} and TbMn$_{2}$O$_{5}$ ~\cite{Hur} and 4.2 ps/m in Cr$_2$O$_3$ ~\cite{Schmid} have been reported. Ba$_{0.5}$Sr$_{1.5}$Zn$_{2}$(Fe$_{1-x}$Al$_{x}$)$_{12}$O$_{22}$ also exhibits substantial MD effects at low temperatures. Figure 3(b) shows $\Delta$$\varepsilon$($B$)/$\varepsilon$(3 T) curves for $x$ = 0.08 below 130 K. The $\varepsilon$, showing a maximum around 0 mT, rapidly decreases within low $B$-ranges ($\mid$$B$$\mid$ $<$ 50 mT), and $B$ = 20 mT is sufficient to induce about 6~\% MD effect. Varying $x$, the MD effects reached as high as 12~\% for $x$ = 0.00 - 0.03 and 15~\% for $x$ = 0.04 - 0.12.
\begin{figure}[tbp]
\centering
\includegraphics[width=0.45\textwidth]{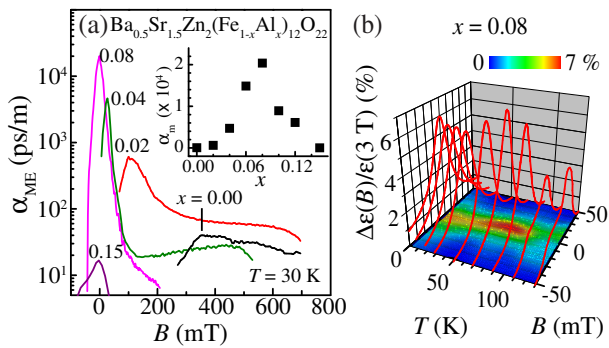}
\caption{(a) $B$-dependence of magnetoelectric susceptibility ($\alpha$$_{\rm ME}$ ) at $T$ = 30 K. The inset shows the maximum values of $\alpha$$_{\rm ME}$ ($\alpha$$_{\rm m}$)  for each $x$. (b) Intensity and 3 D-plots of magnetodielectric effects ($\Delta$$\varepsilon$($B$)/$\varepsilon$(3 T)) of the $x$ = 0.08 sample below 130 K in low $B$ region ($\mid$$B$$\mid$ $<$ 50 mT).}
\label{FIG. 3}
\end{figure}

The observed ferroelectricity induced at extremely low $B$ is likely associated with a heliconical spin ground state. The emergence of spontaneous magnetization along [001] with the increase of $x$ (Fig. 4(a)) supports the hypothesis that the Al-substitution progressively stabilizes the longitudinal heliconical state (Fig. 1(c), left) as reported in Ba$_{2}$Mg$_{2}$Fe$_{12}$O$_{22}$ ~\cite{Ishiwata,Taniguchi}. The estimated conical ordering temperature $T$$_{\rm con}$ is $\sim$20 K for $x$ = 0.02, and it increases up to $\sim$110 K for $x$ = 0.06 and 0.08. In the phase diagram of Fig. 2(c), $B$$_{\rm c,L}$ values for $x$ = 0.06 and 0.08 indeed decrease rather abruptly near their $T$$_{\rm con}\sim$110 K to become less than 10 mT, showing a close link between the heliconical spin state and the extremely low $B$$_{\rm c,L}$. According to the spin current model~\cite{Katsura} that has explained induced $P$ in various spiral- or heli-magnets, $\vec{P}$ $\propto$ $\vec{k}$$_0$ $\times$ [$\vec{\mu}$$_S$ $\times$ $\vec{\mu}$$_L$]. In the present hexaferrite, $\vec{k}$$_0$ is along [001] and thus the model consistently explains why $P$ becomes trivial in the helical or modified helix phases (Fig. 1(b)). Even for the longitudinal conical state (Fig. 1(c), left), $P$ should go to zero as the [$\vec{\mu}$$_S$ $\times$ $\vec{\mu}$$_L$] vector, parallel to the spin-rotation axis of each cone, is along [001]. Under a small in-plane $B$, only the spin-rotation axis of the longitudinal conical state is expected to align easily along the $B$-field to form a transverse conical state (Fig. 1(c), right), which then generates a finite $P$ in the $ab$-plane. Therefore, it is most likely that below $T$$_{\rm con}$ the spin state changes from a longitudinal to a transverse conical type across $B$$_{\rm c,L}$ in the $x$ = 0.06 and 0.08 samples.
\begin{figure}[tbp]
\centering
\includegraphics[width=0.45\textwidth]{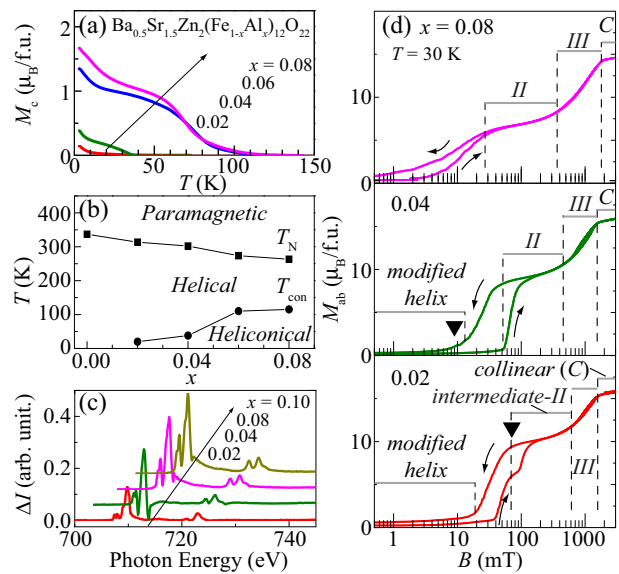}
\caption{(a) The $c$-axis magnetization $M$$_{\rm c}$ measured at zero $B$ after field cooling at $B$ = 3 T.  (b) The magnetic phase evolution with $x$, estimated from $M$$_{\rm c}$($T$) and $M$$_{\rm ab}$($T$) curves. (c) X-ray absorption intensity for several $x$ at the Fe $L$$_{2,3}$-edge after subtraction of the intensity of $x$ = 0.00 ($\Delta$$I$ = -[$I$($\omega$, $x$)-$I$($\omega$, $x$ = 0.00)]). The data for $x$ $\geq$ 0.04 are shifted by every 3.5 eV for clarity. (d) $M$$_{\rm ab}$($B$) curves at $T$ = 30 K. The dashed lines represent the phase boundaries estimated for the down-sweep following the criteria in Ref. ~\cite{Momozawa}. The black triangles represent $B$$_{\rm c,L}$ for $x$ = 0.02 and 0.04. } \label{Fig. 4}
\end{figure}

What can be then the spin state of the FE phases in the low $x$ ($\leq$ 0.04) or possibly above $T$$_{con}$ in the high $x$ ($\geq$ 0.06) samples, in which larger $B$$_{\rm c,L}$ is observed? The in-plane magnetization $M$$_{\rm ab}$ curves (Fig. 4(d)) reveal some clues to this puzzle. According to Ref. ~\cite{Momozawa}, the modified helix, intermediate-I, -II, -III and collinear ferrimagnetic phases coincide with the first, second and third plateau, the steep ascent, and the last plateau regions, respectively. With these criteria, the magnetic phase boundaries can be roughly estimated from Fig. 4(d); for example, the $x$ = 0.02 sample exhibits magnetic phase boundaries similar to BSZFO ~\cite{Kimura03,Momozawa}. However, we notice that the $B$-range for the modified helix phase decreases progressively as $x$ increases to 0.04. For $x$ = 0.08 (also $x$ = 0.06), the first two plateaus of $M$$_{\rm ab}$($B$) no longer exist during the up-sweep, suggesting that the modified helix and intermediate-I phases change into or coexist with the transverse conical state. These observations suggest that the in-plane magnetic anisotropy decreases with increasing $x$, allowing the nontrivial moment along [001] to develop at smaller $B$. Meanwhile, the location of $B$$_{\rm c,L}$ for the $x$ = 0.02 and 0.04 (black triangles) is not anymore coincident with the intermediate-III phase boundary but is well inside the intermediate-II or modified helix phases. Therefore, it is likely that the low $x$-samples also form a kind of transverse conical spin state across their $B$$_{\rm c,L}$, which can coexist with the helical phases, and generate nontrivial $ab$-components of [$\vec{\mu}$$_S$ $\times$ $\vec{\mu}$$_L$] and thus $P$. A previous study of Ba$_{0.4}$Sr$_{1.6}$Zn$_{2}$Fe$_{12}$O$_{22}$ supports this possibility through neutron scattering and magnetic torque measurements~\cite{Perekalina}; near the intermediate-III phase boundary, a cone of easy magnetization, compatible with the transverse conical state in Fig. 1(c), develops and coexists with a basal plane of easy magnetization, compatible with the helical phases in Fig. 1(b).

To understand the role of Al-substitution microscopically, we have performed Fe $L$$_{2,3}$-edge X-ray absorption spectroscopy. The spectra taken for each $x$ were subtracted from that of $x$ = 0.00 to plot the difference spectra as shown in Fig. 4(c). The systematic increase in the intensity of the difference spectra near the octahedral Fe $L$$_{2,3}$-edge supports that Al replaces the octahedral Fe-sites in proportion to $x$. Furthermore, comparative neutron diffraction studies for $x$ = 0.00 and 0.06 produced a good refinement for $x$ = 0.06 when the fraction of Al ions at the octahedral (tetrahedral) sites are 0.9920 (0.0080), directly supporting that Al ions mostly replaces the octahedral Fe sites for $x$ = 0.06. These results support that Al ions in Ba$_{0.5}$Sr$_{1.5}$Zn$_{2}$(Fe$_{1-x}$Al$_{x}$)$_{12}$O$_{22}$ highly prefer to occupy the octahedral Fe sites.

Moreover, magnetic circular dichroism measurements at the Fe $L$$_{2,3}$-edge in the undoped sample revealed a considerable orbital magnetic moment of 0.30 $\pm$ 0.1 $\mu$$_{B}$/f.u. along the $ab$-plane~\cite{Ko}. Note that ionic Fe$^{3+}$ has a half-filled $d^5$ configuration and its total angular momentum $L$ = 0. Thus the orbital moment is expected to vanish. If the Fe$^{3+}$ ion, however, is not at the center of an octahedron, i.e., off-centered, then the hybridization with O 2$p$ becomes anisotropic and contributes a non-vanishing orbital moment, as observed in GaFeO$_{3}$ ~\cite{Kim}. According to our structural refinements as well as those in Ref. ~\cite{Momozawa}, the Fe$^{3+}$ ions in the green octahedra (Fig. 1(a)) are indeed off-centered, being shifted along the $c$-axis, which results in the nontrivial orbital moment and the easy spin axis in the $ab$-plane. Based on the octahedral site preference of the Al$^{3+}$ ions, it is likely that the main role of the Al-substitution is to reduce the magnetic anisotropy energy, through either dilution of the off-centered Fe$^{3+}$ sites or reduction of their off-centering deformation. Indeed, subsequent measurements for a $x$ = 0.08 sample showed decrease of the in-plane orbital magnetic moment by $\sim$20 \% ~\cite{Ko}.  As a consequence, the magnetic moment can tilt away from the $ab$-plane more easily with Al-substitution, resulting in reduced $B$$_{\rm c,L}$ at low $x$, and the heliconical spin state seems to emerge for $x$ = 0.06 - 0.08 with reduced in-plane anisotropy. Therefore, decrease of the in-plane orbital moment with the Al-substitution seems to be an essential process for lowering magnetic anisotropy and thus improving the low field ME response ~\cite{highdoping}.

The mechanism uncovered here is also applicable to Ba$_{2}$Mg$_{2}$Fe$_{12}$O$_{22}$ ~\cite{Ishiwata,Taniguchi}, in which Mg$^{2+}$ ions occupy both tetrahedral and octahedral Fe sites in an arbitrary fraction. Here, the partial occupation of Mg$^{2+}$ in the octahedral Fe-sites seems to also reduce the magnetic anisotropy~\cite{Ishiwata2}, via the decrease of the orbital moment. In comparison, Al-substitution in BSZFO seems to be more effective for systematic control of magnetic anisotropy and $B$$_{\rm c,L}$ as it can directly control the orbital moment in the system.

Finally, it is noteworthy that the helical spin ordering temperatures of Al-substituted BSZFO remain relatively high (Fig. 4(b)); for instance, $T$$_{\rm N}$ = 337 K at $x$ = 0.00 and 263 K at $x$ = 0.08. Although the FE phase boundaries could not be determined above 220 K due to electrical leakage, we expect that the low $B$-induced FE switching will be observable up to $T$$_{\rm N}$ upon further optimization of the resistivity. In our recent work on a heat-treated BSZFO sample, $B$$_{\rm c,L}$ could be determined up to $T$$_{\rm N}$ = 315 K due to dramatic increase of resistivity ~\cite{Chai}. Therefore, there is high hope that the Al-substituted BSZFO also exhibits low $B$-induced ferroelectricity even near room temperature.

In summary, we have shown that Al-substitution into hexaferrites provides an unprecedented opportunity to decrease effectively the critical magnetic field for inducing ferroelectricity as well as to obtain gigantic low field magnetoelectricity through the control of orbital moment. Upon optimization of spin ordering temperature and electrical resistivity, the helimagnetic insulators may provide a pathway to the realization of novel magnetoelectric devices at room temperature under low magnetic fields.

This work is supported by NRL (M10600000238), GPP (K20702020014-07E0200-01410), and  Basic Science Research (2009-0083512) programs. JHP is supported by National Creative Initiative Center and WCU program (R31-2008-000-10059-0) and JHC by KOSEF (20090059529) and BAERI programs. SHC is supported by Seoul R\&BD (10543) and Seoul Science Fellowship.


\begin{thebibliography}{24}
\bibitem{Kimura01} T. Kimura {\it et al.}, Nature (London) {\bf 426}, 55 (2003).
\bibitem{Hur} N. Hur {\it et al.}, Nature (London) {\bf
429}, 392 (2004).
\bibitem{Cheong} S.-W. Cheong and M. Mostovoy, Nature Mater. {\bf 6}, 13 (2007).
\bibitem{Ramesh} R. Ramesh and N. A. Spaldin, Nature Mater. {\bf 6}, 21 (2007).
\bibitem{Nan} C.-W. Nan {\it et al.}, J. Appl. Phys. {\bf 103}, 031101 (2008).
\bibitem{Katsura} H. Katsura, N. Nagaosa, and A. V. Balatsky, Phys. Rev. Lett. {\bf 95}, 057205 (2005).
\bibitem{Mostovoy} M. Mostovoy, Phys. Rev. Lett. {\bf 96}, 067601 (2006).
\bibitem{Sergienko} I. A. Sergienko and E. Dagotto, Phys. Rev. B {\bf 73}, 094434 (2006).
\bibitem{Kenzelmann} M. Kenzelmann {\it et al.}, Phys. Rev. Lett. {\bf 95}, 087206 (2005).
\bibitem{Chapon} L. C. Chapon {\it et al.}, Phys. Rev. Lett. {\bf 93}, 177402 (2004).
\bibitem{Kimura02} T. Kimura {\it et al.}, Nature Mater. {\bf 7}, 291 (2008).
\bibitem{Kimura03} T. Kimura, G. Lawes, and A. P. Ramirez, Phys. Rev. Lett. {\bf 94}, 137201 (2005).
\bibitem{Eerenstien} W. Eerensiein {\it et al.}, Nature Mater. {\bf 6}, 348 (2007).
\bibitem{Dong} S. Dong, J. Zhai, J. Li, and D. Viehland, Appl. Phys. Lett. {\bf 89}, 252904 (2006).
\bibitem{Ishiwata} S. Ishiwata {\it et al.}, Science {\bf 319}, 1643 (2008).
\bibitem{Taniguchi} K. Taniguchi {\it et al.}, Appl. Phys. Express {\bf 1}, 031301 (2008).
\bibitem{Momozawa} N. Momozawa and Y. Yamaguchi, J. Phys. Soc. Jpn. {\bf 62}, 1292 (1993).
\bibitem{Schmid} H. Schmid, in \emph{Introduction to Complex Mediums for Optic and Electromagnetics}, ed. by W. S. Weiglhofer and A. Lakhtakia, (SPIE Press, Bellingham, 2003), pp.167.
\bibitem{Perekalina} T. M. Perekalina {\it et al.}, Sov. Phys. JETP {\bf 25}, 266 (1967).
\bibitem{Ko} W.-S. Noh {\it et al.}, to be published.
\bibitem{Kim} J.-Y. Kim, T. Y. Koo, and J.-H. Park, Phys. Rev. Lett. {\bf 96}, 047205 (2006).
\bibitem{highdoping} For $x$ $>$ 0.08, $P$ already exists even at zero $B$, resulting in less change of $P$ with $B$, and $P$$_{\rm max}$ decreases systematically, both of which can explain the decrease of ME suceptibility.
\bibitem{Ishiwata2} S. Ishiwata {\it et al.}, Phys. Rev. B {\bf 79}, 180408(R) (2009).
\bibitem{Chai} Y. S. Chai {\it et al.}, New. J. Phys. {\bf 11}, 073030 (2009).
\end{thebibliography}
\end{document}